\documentclass[prl,twocolumn,showpacs,superscriptaddress]{revtex4-1}
\usepackage{graphicx,helvet}
\usepackage{amsmath}
\usepackage{amsfonts}
\usepackage{amssymb}
\usepackage{wasysym}
\usepackage{dsfont}
\usepackage{color,soul}

\begin{document}

\title{Synchronization Dynamics in the Presence of Coupling Delays and Phase Shifts}

\author{David J. J\"org}
\affiliation{Max Planck Institute for the Physics of Complex Systems, N\"othnitzer Str. 38, 01187 Dresden, Germany}

\author{Luis G. Morelli}
\affiliation{Departamento de F\'{\i}sica, FCEyN UBA and IFIBA, Conicet; Pabell\'on 1, Ciudad Universitaria, 1428 Buenos Aires, Argentina}

\author{Sa\'ul Ares}
\affiliation{Grupo Interdisciplinar de Sistemas Complejos (GISC), and Logic of Genomic Systems Laboratory, Centro Nacional de Biotecnolog\'ia --- CSIC, Calle Darwin 3, 28049 Madrid, Spain}

\author{Frank J\"ulicher}	 \email{julicher@pks.mpg.de}
\affiliation{Max Planck Institute for the Physics of Complex Systems, N\"othnitzer Str. 38, 01187 Dresden, Germany}

\date{\today}

\begin{abstract} 
\noindent 
In systems of coupled oscillators, the effects of complex signaling can be captured by time delays and phase shifts.
Here, we show how time delays and phase shifts lead to different oscillator dynamics and how synchronization rates can be regulated by substituting time delays by phase shifts at constant collective frequency.
For spatially extended systems with time delays, we show that fastest synchronization can occur for intermediate wavelengths, giving rise to novel synchronization scenarios.
\end{abstract}


\maketitle

\noindent It has been known for over three-hundred years that interacting dynamic oscillators generally tend to synchronize, even if interactions are weak \cite{StrogatzStewart}.
This synchronization occurs robustly and independent of the details of the interaction mechanism.
A simple model for the generic features of synchronization is the Kuramoto model \cite{Kuramoto,Acebron}.
It describes the phase dynamics of instantaneously coupled oscillators using a periodic coupling function.
The instantaneous frequency of a given oscillator is influenced by the phase received from other oscillators.
In general, coupling tends to keep the phase difference between the oscillator and the received signal at a constant value~$\alpha$.
For $\alpha=0$, two oscillators tend to synchronize, for $\alpha=\pi$, they tend to lock in anti-phase.

In many systems, signaling processes are complex and signaling times cannot be ignored. 
For example, in biological systems, dynamic oscillators are often coupled via complex molecular signaling processes~\cite{Lewis,GarciaOjalvo}.
If the processes involved take a time comparable to the oscillation cycle, these time delays in the coupling can play a significant role for the dynamics of the system and the properties of synchronized states~\cite{Yeung}.
In principle, effects of signaling times could be captured either by modifying the phase shifts~$\alpha$ or by introducing an explicit time delay~$\tau$~\cite{Izhikevich}.
The effects of time delays have been studied extensively.
In particular, coupling delays can lead to multistability of synchronized states and affect their collective frequency \cite{Yeung,Ares,Herrgen}.
While often considered as an undesired but inevitable feature, constructive roles of coupling delays on synchronization have been reported \cite{Dhamala,Rosenblum}.
It has been shown for systems with both phase shifts and time delays, that in the synchronized state, time delays effectively induce an additional effective phase shift between coupled oscillators~\cite{Yeung}, suggesting that phase shifts alone may capture the essential effects of delays.
This raises the question whether phase shifts and time delays play a similar role in networks of coupled oscillators.
In this Letter, we show that gradually substituting time delays by phase shifts, keeping the collective frequency constant, there exists a specific combination of time delay and phase shift for which the rate of synchronization is fastest.
This applies both to globally coupled oscillators as well as different coupling topologies.
In spatially extended systems, substituting time delays by phase shifts can regulate the length scale at which synchronization is fastest.
Our results demonstrate how the phase shift~$\alpha$ and the delay~$\tau$ account for different physical effects of complex oscillator coupling.

We obtain our results using a Kuramoto model for a network of identical coupled oscillators, which takes into account the time delay~$\tau$ and phase shift~$\alpha$~\cite{Yeung,SakaguchiKuramoto},
\begin{align}
\frac{d}{dt}\theta_i(t) = \omega + \frac{K}{\rho_i} \sum_{j=1}^N a_{ij} \Gamma(\theta_j(t-\tau)-\theta_i(t)-\alpha) \ .
\label{cpo.model}
\end{align}
Here, $\theta_i(t)$ is the phase of oscillator $i$, $N$ is the total number of oscillators, $\omega $ is the intrinsic frequency of the oscillators. 
Oscillator coupling of strength $K$ is described by the $2\pi$-periodic function $\Gamma(\vartheta)$.
The adjacency matrix~$a_{ij}$ with $a_{ij} \geq 0$ defines the coupling topology and $\rho_i=\sum_j a_{ij}$ is the total weight of links of oscillator~$i$.

Because of the normalization of the coupling strength by $\rho_i$ in Eq.~(\ref{cpo.model}), in-phase synchronized states with $\theta_i(t) = \Omega t$ always exist.
The collective frequency~$\Omega$ obeys the equation \cite{SchusterWagner,Yeung}
\begin{align}
\Omega &= \omega + K \Gamma(-\Omega\tau-\alpha) \ . \label{Omega}
\end{align}
The collective frequency thus depends on $\alpha$ and $\tau$. For $\tau > 0$, several synchronized states with different collective frequency can coexist.

By simultaneously changing $\alpha$ and $\tau$, it is possible to keep the collective frequency constant.
For any synchronized state with collective frequency $\Omega=\Omega_0$ obeying Eq.~(\ref{Omega}), the transformation $\tau \to \tau'$, $\alpha \to \alpha + \Omega_0(\tau-\tau')$ preserves the existence of the synchronized state with~$\Omega=\Omega_0$.
This transformation implies that for a given value of $\Omega_0\tau+\alpha$, there is a one-parameter family of systems in the $(\tau,\alpha)$-plane that can exhibit the same collective frequency.
These systems can be parameterized as
\begin{align}
	\alpha(\tau)  &= \psi - \Omega_0 \tau \ , \label{alpha}
\end{align}
where $\psi$ is a constant that sets the collective frequency~$\Omega_0$ as
\begin{align}
	\Omega_0 = \omega + K \Gamma(-\psi) \ .
\end{align}

Varying~$\tau$ and using the phase shift $\alpha(\tau)$, the collective frequency does not change.
However, here we will show that the synchronization dynamics does change.
To study the dependence of the synchronization dynamics on time delays and phase shifts at constant frequency, we introduce a small perturbation to the synchronized state and determine its exponential relaxation rate $r_0$. We compute $r_0$ below both analytically and from numerical simulations. Numerically, $r_0$ can be determined from the exponential relaxation time of the perturbation to perfect synchrony, monitored by the Kuramoto order parameter \cite{Wetzel2012} \footnote{We prepare the system in the synchronized state and introduce small random perturbations to all phases. We let this perturbed state relax, and measure the Kuramoto order parameter $Z(t) = N^{-1} | \sum_i \mathrm{e}^{\mathrm{i}\theta_i(t)} |$ as it approaches one. Relaxation becomes exponential for large times. We determine the relaxation rate from a fit to this exponential.}.
\begin{figure}[t]
\begin{center}
\includegraphics[width=8.6cm]{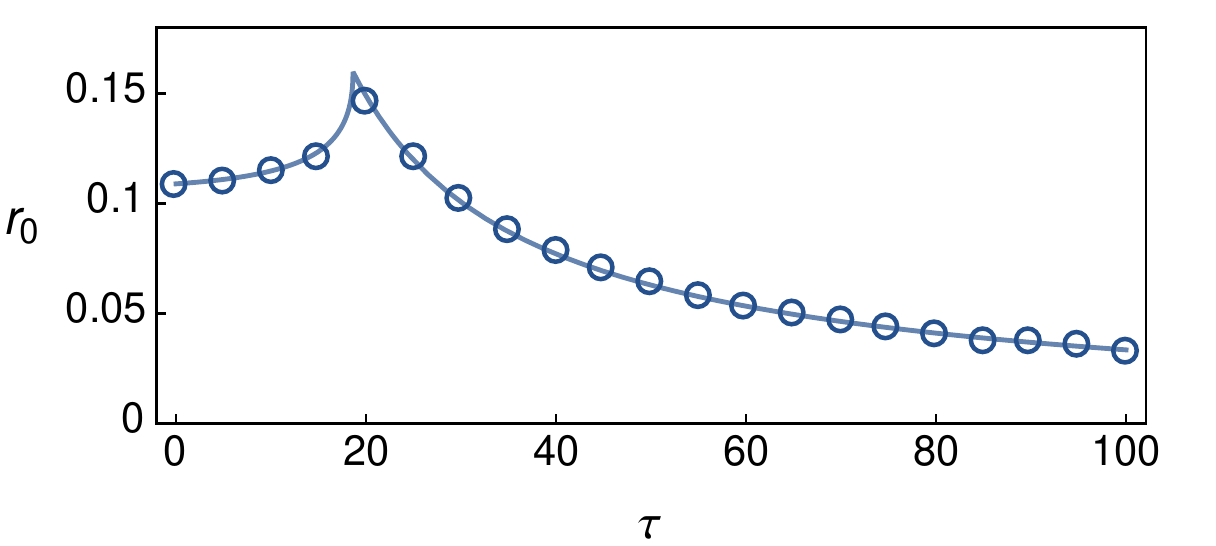}
\caption{Synchronization rate $r_0$ as a function of the coupling delay~$\tau$ for a globally coupled system. Circles: numerical simulations of Eqs.~(\ref{cpo.model}) and (\ref{alpha}); line: Eq.~(\ref{sol.W.alt}). $\Gamma(\vartheta)=\sin \vartheta$, $N=40$, $\psi=5.5$, $\omega=1$, and $K=0.15$. Collective frequency: $\Omega_0 = 1.11$.}
\label{fig.mfop}
\end{center}
\end{figure}

As a first example, we consider globally coupled oscillators with $\Gamma(\vartheta)=\sin\vartheta$.
Fig.~\ref{fig.mfop} displays the relaxation rate of the system as a function of the time delay~$\tau$, obtained by numerical integration of Eq.~(\ref{cpo.model}) (circles).
This result shows that in this system, there is a characteristic value of the coupling delay for which the synchronization rate is maximal. The analytical solution for the synchronization rate (solid line), derived below, displays a characteristic cusp where this maximum is attained.

As a second example, we consider a system with nearest-neighbor coupling in one dimension with periodic boundary conditions, see Fig.~\ref{fig.rtau}.
In this case, as shown below, we find that spatial Fourier modes of the oscillator lattice relax independently, each with a relaxation rate $r_0(k)$ that depends on the wavevector $k=2 \pi p/N$ of the Fourier mode.
There exists a discrete set of wavevectors for which $p \in \{-\frac{N}{2},-\frac{N}{2}+1,\hdots,\frac{N}{2}-1\}$, where $N$, considered to be even, is the system size.
Note that because of the delays, there is in fact a discrete set of relaxation rates~$r_n(k)$ for a given wavevector.
However, synchronization is governed by the slowest rate~$r_0$.
The relaxation rate~$r_0$ of long-wavelength modes, $|k|<\pi/2$, decreases with increasing wavelength and time delay, see Fig.~\ref{fig.rtau} (dashed red lines).
Fourier modes with short wavelengths, $|k|>\pi/2$, display a cusp-like maximum, a behavior that was already observed in a different system as shown in Fig.~\ref{fig.mfop}. The delay~$\tau$ corresponding to this maximum and the corresponding relaxation rate~$r_0$ depend on the wavevector~$k$.

\begin{figure}[t]
\begin{center}
\includegraphics[width=8.6cm]{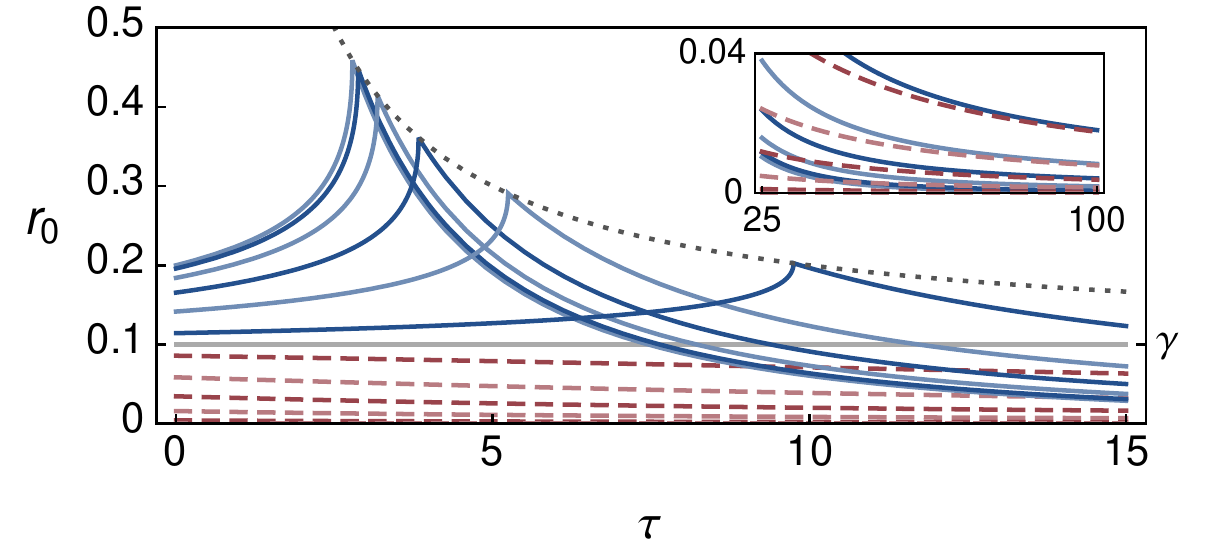}
\caption{Synchronization rates~$r_0$ for a nearest-neighbor coupled system and periodic boundary conditions in one dimension with $N=22$~oscillators and $\gamma=0.1$. 
At $\tau=0$ the curves are ordered from bottom to top in direction of increasing wavevector $k=2\pi p/N$. Dashed red lines: $p=1,\hdots,5$, solid blue lines: $p=6,\hdots,11$.
Adjacent curves have different brightness for visual clarity. Dotted line: envelope for the maxima $\gamma+1/\tau$.
Inset: same curves for large time delays.
}
\label{fig.rtau}
\end{center}
\end{figure}

In order to better understand these examples and to obtain basic insights in the behavior of the large class of systems described by Eq.~(\ref{cpo.model}), we perform a general study of the relaxation rate as a function of coupling delay and phase shift.
We consider the linear dynamics near the synchronized state, $\theta_i(t) = \Omega_0 t + \varepsilon \xi_i(t)$ with $\varepsilon \ll 1$.
For simplicity, we focus on coupling topologies, for which the normalized adjacency matrix~$b_{ij} = a_{ij}/\rho_i$ is symmetric and $b_{ii}=0$ for all~$i$.
The two examples introduced above fall within this class of systems.
Following~\cite{EarlStrogatz}, the in-phase synchronized state of Eq.~(\ref{cpo.model}) is stable if and only if $\gamma > 0$, where $\gamma \equiv K \frac{d}{d\vartheta} \Gamma(\vartheta) |_{\vartheta=-\psi}$. We  only consider these stable cases.

To first order in $\varepsilon$, the time evolution of the perturbation is given by
\begin{align}
\frac{d}{dt}\xi_i(t) = \gamma \sum_{j=1}^N b_{ij} [ \xi_j(t-\tau)-\xi_i(t) ] \ .
\label{gov.pb}
\end{align}
We introduce the collective relaxation modes $\phi_i(t) \equiv \sum_{j} d^{-1}_{ij} \xi_j(t)$, where $d_{ij}$ is defined by $\sum_{jk} d^{-1}_{ij} b_{jk} d_{kl} = u_i \delta_{il}$ and  $u_i$ are the $N$~eigenvalues of the matrix $b_{ij}$. The eigenvalues $u_i$ are real and obey $|u_i|\leq 1$ \cite{EarlStrogatz}.
To compute the relaxation of collective modes, we take the time derivative of $\phi_i(t)$ and use Eq.~(\ref{gov.pb}) to replace $\xi_j(t)$.
Inserting the identity $\delta_{ij} = \sum_k d_{ik} d_{kj}^{-1}$ enables to express the result in terms of the collective modes $\phi_i$ and the eigenvalues $u_i$.
The collective modes relax independently according to
\begin{align}
\frac{d}{dt}\phi_i(t) = \gamma [u_i \phi_i(t-\tau) - \phi_i(t)] \ .
\label{dp.te}
\end{align}
The ansatz $\phi(t) = \mathrm{e}^{-\lambda t}$ yields the  characteristic equation for the relaxation rates $\lambda$ of the collective mode~$\phi$ \cite{MacDonald,Atay},
\begin{align}
\gamma-\lambda = \gamma u \mathrm{e}^{\lambda \tau} \ ,
\label{te.lambda}
\end{align}
where we have dropped the index $i$ for notational simplicity.
Solutions to Eq.~(\ref{te.lambda}) can be expressed in terms of the Lambert $W$ function \cite{Amann}, 
defined by the relation $W(z) \mathrm{e}^{W(z)} = z$ for $z \in \mathds{C}$. This function has discrete branches $W_n(z)$ separated by branch cuts, where $n$ is the branch index~\cite{Corless}.
Each branch~$n$ of $W$ corresponds to one relaxation rate~$r_n=\operatorname{Re} \lambda_n$.
Note that our sign convention for $\lambda$ implies that for stable states, all $r_n$ are positive.
Here, we focus on the slowest relaxation rate~$r_0$ for a given collective mode~$\phi$, which corresponds to the long time behavior of~$\phi$.
Solving Eq.~(\ref{te.lambda}) for $\lambda$, the solution~$\lambda_0$ is
\begin{align}
\lambda_0 &= \gamma - \frac{1}{\tau} W_0 \! \left( z_\tau \right) \ , \label{sol.W.alt}
\end{align}
where $z_\tau \equiv  u \gamma\tau \mathrm{e}^{\gamma\tau}$, since the principal branch $W_0$ has the property $\operatorname{Re}W_0 \geq \operatorname{Re}W_n$~\cite{ShinozakiMori}. The dependence of $\lambda_0$ on the coupling delay~$\tau$ thus depends on the properties of the principal branch~$W_0$ of the Lambert function.

To discuss the properties of the slowest relaxation rates~$r_0$, we consider separately collective modes with $u>0$ and $u<0$.
In nearest-neighbor coupled systems, collective modes with $u>0$ are the Fourier modes with long wavelengths, as shown below.
For these modes, $r_0$ decreases monotonically and converges to zero for $\tau\to\infty$ (dashed lines in Fig.~\ref{fig.rtau}).
This can be shown using Eq.~(\ref{te.lambda}) and writing
\begin{align}
\gamma-r &= u \gamma \mathrm{e}^{\tau r} \cos(\tau \nu) \ , \label{r} \\
-\nu &= u \gamma \mathrm{e}^{\tau r} \sin(\tau \nu) \ ,
\label{s}
\end{align}
where $\nu=\operatorname{Im} \lambda$.
The smallest value of $r=r_0$ corresponds to $\cos(\tau \nu_0)=1$. From Eq.~(\ref{s}), it then follows that $\nu_0=0$.
Using Eq.~(\ref{te.lambda}), we find
\begin{align}
\frac{dr_0}{d\tau} = - \frac{r_0}{\tau + (\gamma-r_0)^{-1}} < 0 \ .
\label{r.monotonicity}
\end{align}
Furthermore, Eq.~(\ref{te.lambda}) implies that 
$\tau= \ln ( u [1-{r_0}/{\gamma} ] )/r_0$, which reveals that $r_0 \to 0$ corresponds to $\tau\to\infty$. Therefore, $r_0$ vanishes for large $\tau$.
Hence, the collective modes corresponding to $u>0$ become stationary for large time delay.
Eq.~(\ref{dp.te}) furthermore implies that given two eigenvalues $u_1 \geq u_2$, the respective exponents satisfy $r_0^{(1)} \geq r_0^{(2)}$ for all $\tau$ by an argument similar to the one leading to Eq.~(\ref{r.monotonicity}). In Fig.~\ref{fig.rtau}, this is illustrated by the fact that the dashed lines never cross.

In the case of collective modes with $u<0$ in Eq.~(\ref{te.lambda}), it can be shown that $r_0$ displays a cusp at $\tau=\tau^*$, where
\begin{align}
\tau^* \equiv \frac{1}{\gamma} W_0\left(\frac{\mathrm{e}^{-1}}{|u|}  \right) \ .
\label{minimum}
\end{align}
At $\tau=\tau^*$, $r_0$ is not analytic and its first derivative has a jump which stems from the definition of the principal branch $W_0$ of the Lambert function.
We now show that $\frac{dr_0}{d\tau}$ has opposite sign in the two regions $\tau \leq \tau^*$ and $\tau>\tau^*$. Therefore, the cusp is located at the maximum of $r_0$, as suggested by Figs.~\ref{fig.mfop} and \ref{fig.rtau}. Differentiating Eq.~(\ref{sol.W.alt}) with respect to $\tau$ and using the defining relation of the Lambert $W$ function to compute its derivative, we obtain, with $W_0(z_\tau)\equiv U+\mathrm{i}V$,
\begin{align}
\frac{d r_0}{d\tau} = \frac{1}{\tau^2} \frac{({U}-\gamma\tau)({U}+{U}^2 + {V}^2)-{V}^2}{(1+{U})^2 + {V}^2} \ .
\label{real.drv}
\end{align}
For $\tau \leq \tau^*$, we find $V=0$.
This follows from the properties of the principal branch $W_0$, in particular $\operatorname{Im} W_0(z) = 0$ for $z \geq -\mathrm{e}^{-1}$.
Since ${U} \in [-1,0]$, Eq.~(\ref{real.drv}) implies $\frac{d}{d\tau} r_0 \geq 0$ for $\tau \leq \tau^*$.

For $\tau>\tau^*$, we show that $\frac{dr_0}{d\tau}$ is negative. In Eq.~(\ref{real.drv}), the factor ${U}-\gamma\tau$ is negative. This can be seen by taking the real part of Eq.~(\ref{sol.W.alt}) and using the fact that $r_0>0$.
Furthermore, $V\neq 0$ in this region. The factor ${U}+{U}^2 + {V}^2$ is positive: For $\tau>\tau^*$, we have ${U} = - {V} \cot {V}$ and the numerator of the second factor in (\ref{real.drv}) can be rewritten as
\begin{align}
{U}+{U}^2 + {V}^2
&= \frac{{V}}{(\sin {V})^2} \left( {V} - \frac{\sin(2 {V})}{2} \right) \ .
\end{align}
Since ${V} \in [0,\pi]$, the above expression is positive. Altogether, we conclude that for $u<0$, $\frac{d}{d\tau} r_0 \leq 0$ for $\tau >\tau^*$. The corresponding collective modes therefore resynchronize slower as the time delay increases.
The maximal resynchronization rate~$r_0^*$ at $\tau=\tau^*$ is given by $r_0^* = \gamma + 1/\tau^*$.

The behavior of $r_0$ in the limit of large $\tau$ can be obtained from an expansion of $r_0$ in powers of $\tau^{-1}$, $r_0 = -\ln |u|/\tau + \mathcal{O}(\tau^{-2})$, which reveals that for collective modes with eigenvalues $u$ and $-u$, the synchronization rate $r_0$ approaches the same asymptotic behavior for large $\tau$. 
The inset of Fig.~\ref{fig.rtau} reflects this property for the case of nearest-neighbor coupling.

\begin{figure}[t]
\begin{center}
\includegraphics[width=8.5cm]{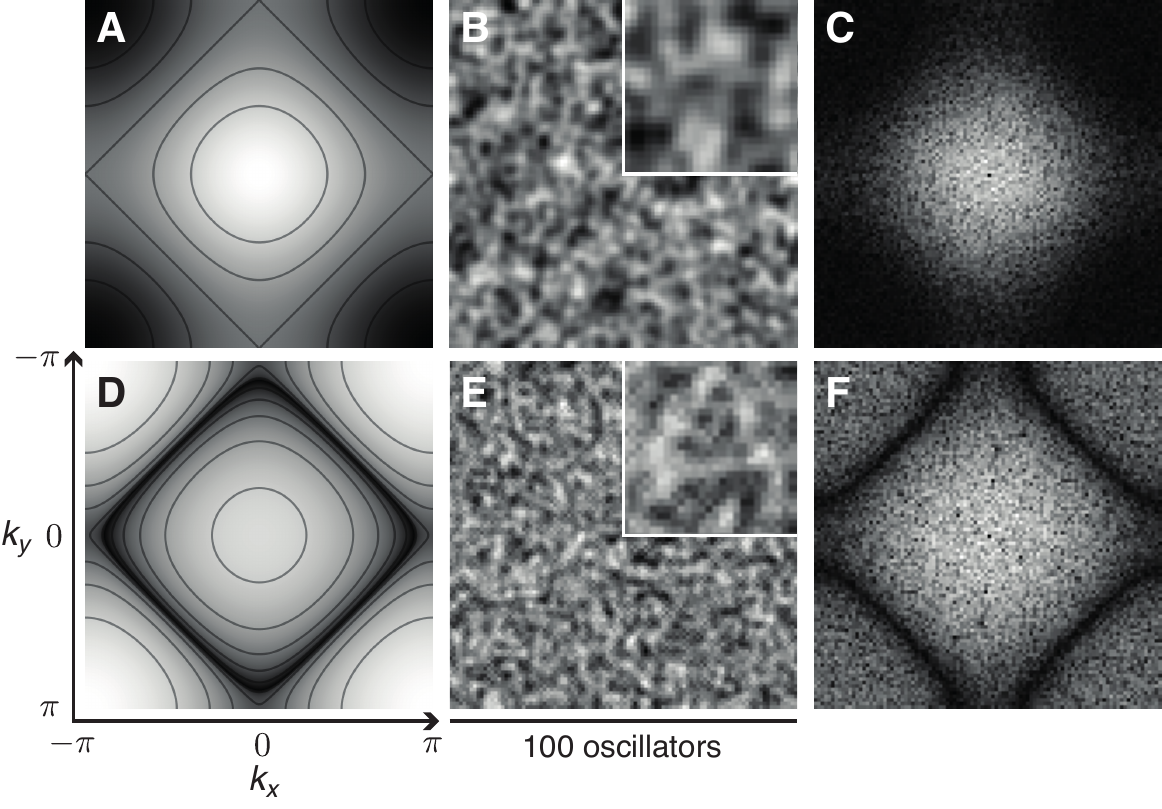}
\caption{Synchronization of oscillators with nearest-neighbor coupling in two dimensions for no coupling delay (A--C) and finite delay (D--F) but same collective frequency.
(A,D) $r_0$ as a function of wavevector
for a regular 2D square lattice. 
Bright colors correspond to small, dark colors to large values. 
Axes scaling equal in both panels. 
(B,E) Simulation snapshots of Eqs.~(\ref{cpo.model}) and (\ref{alpha})
at time $t=24$ with nearest-neighbor sine coupling on a $100\times100$ lattice with periodic boundary conditions. 
Intensity indicates relative values of the sine of the corresponding oscillator's phase.
Initial conditions are the synchronized state, perturbed by phases randomly chosen in $[-0.4\pi,0.4\pi]$.
Insets: $2\times$ magnifications.
(C,F) Logarithmic power spectra of images~B,E.
Axes as in panels~A,D.
Parameters:
$\psi=6$, $\omega=1$, $K=0.2$, $\alpha$ given by Eq.~(\ref{alpha}), and
 $\tau=0$ (A--C) or $\tau=10$ (D--F).
Collective frequency:~$\Omega = 1.06$.
}
\label{fig.bdries}
\end{center}
\end{figure}

The mode structure shown in Fig.~\ref{fig.rtau} can be understood as follows. For nearest-neighbor coupling in $d$~dimensions with periodic boundary conditions, the collective modes are Fourier modes of the linear perturbations $\xi$ satisfying Eq.~(\ref{gov.pb}):
\begin{align}
\phi_{k_1,\hdots,k_d}(t) =
\sum_{j_1=0}^{N_1-1} \hdots \sum_{j_d=0}^{N_d-1} \mathrm{e}^{-\mathrm{i} \sum \limits_{n=1}^d k_n j_n} \xi_{j_1,\hdots,j_d}(t) \ .
\label{fourier}
\end{align}
These collective modes satisfy Eq.~(\ref{dp.te}) with eigenvalues
\begin{align}
u_{k_1,\hdots,k_d} =  \sum_{n=1}^d \cos k_n \ ,
\label{hd.evs}
\end{align}
where $k_n=2\pi p_n/N_n$ with $p_n \in \{-\frac{N_n}{2},-\frac{N_n}{2}+1,\hdots,\frac{N_n}{2}-1 \}$, and $N_n$ is the size of the system in $n$-direction ($n=1,\hdots,d$). 
Hence, the eigenvalues~$u_{k_1,\hdots,k_d}$ refer to the respective Fourier modes with wavevector $(k_1,\hdots,k_d)$.
The slowest relaxation rate~$r_0$ for this Fourier mode can be obtained taking the real part of $\lambda_0$, given by Eq.~(\ref{sol.W.alt}) with $u=u_{k_1,\hdots,k_d}$.

We illustrate this result for the case of a two-dimensional system, see Fig.~\ref{fig.bdries}.
The synchronization rate $r_0$ is displayed 
as a function of the wavevector~$(k_x,k_y)$ in Fig.~\ref{fig.bdries}A and \ref{fig.bdries}D 
for two systems with no delay and finite delay~$\tau$, respectively, and $\alpha$ chosen according to Eq.~(\ref{alpha}) to impose the same collective frequency. 
For no delay, Eq.~(\ref{sol.W.alt}) leads to the classical scenario where short-wavelength collective modes decay quickly while long-wavelength modes decay slowly (dark corners in Fig.~\ref{fig.bdries}A).
Interestingly, for long delays collective modes decay fastest at intermediate wavelengths (Fig.~\ref{fig.rtau} and dark diamond in Fig.~\ref{fig.bdries}D). 
This remarkable behavior is confirmed by full simulations of Eq.~(\ref{cpo.model}), Figs.~\ref{fig.bdries}B,C,E and F.
The inset of Fig.~\ref{fig.bdries}E shows partially synchronized clusters on intermediate length scales with persisting phase differences on nearest-neighbor scale.
This behavior reflects the fact that the curves for short-wavelength collective modes in Fig.~\ref{fig.rtau} reverse their ordering as $\tau$ increases.
A similar mode reversal has been observed in small systems of chaotic oscillators as a function of the coupling strength~\cite{Heagy}.

The behavior of a globally coupled system, Fig.~\ref{fig.mfop}, can be understood as follows. The normalized adjacency matrix is given by $b_{ij} = (1-\delta_{ij})(N-1)^{-1}$. 
The largest eigenvalue, $u=1$, corresponds to the neutrally stable global phase shift. 
All other collective modes have eigenvalue $u=(1-N)^{-1}$.
These modes therefore exhibit the same synchronization rate whose $\tau$-dependence is non-monotonic. According to Eq.~(\ref{minimum}), the maximal synchronization rate of a system with global coupling is located at $\tau^*=  W_0(\mathrm{e}^{-1}[N-1])/\gamma$ and depends on the system size and properties of the coupling.

In this work we have shown how coupling delays and phase shifts play a different role in regulating synchronization in systems of coupled phase oscillators.
Our results show that synchronization rates can exhibit maxima as a function of time delay when the collective frequency is kept constant by adjusting phase shifts.
Interestingly, in spatially extended systems with time delays, the relaxation rate does not always decrease with increasing wavelength but intermediate wavelengths may relax faster than short ones, giving rise to novel relaxation scenarios.
Phase shifts alone cannot give rise to this behavior.

Fast synchronization improves the resilience of the synchronized state in the presence of fluctuations or diversity~\cite{Pikovsky,Manrubia,RiedelKruse}. 
Here we have considered identical oscillators, but in natural systems diversity of oscillators can introduce a distribution of frequencies.
For a narrow frequency distribution, we have confirmed numerically (data not shown) that a maximum of the synchronization rate still occurs for a non zero coupling delay.
If the function of a system demands the collective frequency to be in a specific small range, the possibility  to regulate synchronization rates using phase shifts and time delays at constant frequency might be important.
Examples for such systems are the core pacemaker of the circadian clock, regulating metabolism in higher organisms with a period of about 24 hours \cite{Zhang,Schibler,Gonze2}, the segmentation clock of vertebrates \cite{Oates}, whose collective frequency determines the length of body segments \cite{Cooke,Morelli,Herrgen,Ares}, and engineered systems of coupled lasers or electronic oscillators \cite{Kozyreff,Tousi}.
Our work shows that together with phase shifts, coupling delays can play an important role 
for the regulation of dynamic behaviors and the resilience of synchronized oscillator networks.

We thank Lucas Wetzel for many fruitful discussions and Douglas B.~Staple for critical comments on the manuscript. We thank Andy Oates and members of his lab for inspiring discussions about biological oscillators. LGM acknowledges funding from PICT 2012-1954 and the hospitality of the Biological Physics Division at MPI--PKS. SA acknowledges funding from the Spanish Ministry of Economy and Competitiveness (MINECO)  through grant PHYSDEV (FIS2012-32349), and from CSIC through the Junta para la Ampliaci\'on de Estudios program (JAEDOC014, 2010 call) co-funded by the European Social Fund.


\begin{thebibliography}{33}%
\makeatletter
\providecommand \@ifxundefined [1]{%
 \@ifx{#1\undefined}
}%
\providecommand \@ifnum [1]{%
 \ifnum #1\expandafter \@firstoftwo
 \else \expandafter \@secondoftwo
 \fi
}%
\providecommand \@ifx [1]{%
 \ifx #1\expandafter \@firstoftwo
 \else \expandafter \@secondoftwo
 \fi
}%
\providecommand \natexlab [1]{#1}%
\providecommand \enquote  [1]{``#1''}%
\providecommand \bibnamefont  [1]{#1}%
\providecommand \bibfnamefont [1]{#1}%
\providecommand \citenamefont [1]{#1}%
\providecommand \href@noop [0]{\@secondoftwo}%
\providecommand \href [0]{\begingroup \@sanitize@url \@href}%
\providecommand \@href[1]{\@@startlink{#1}\@@href}%
\providecommand \@@href[1]{\endgroup#1\@@endlink}%
\providecommand \@sanitize@url [0]{\catcode `\\12\catcode `\$12\catcode
  `\&12\catcode `\#12\catcode `\^12\catcode `\_12\catcode `\%12\relax}%
\providecommand \@@startlink[1]{}%
\providecommand \@@endlink[0]{}%
\providecommand \url  [0]{\begingroup\@sanitize@url \@url }%
\providecommand \@url [1]{\endgroup\@href {#1}{\urlprefix }}%
\providecommand \urlprefix  [0]{URL }%
\providecommand \Eprint [0]{\href }%
\providecommand \doibase [0]{http://dx.doi.org/}%
\providecommand \selectlanguage [0]{\@gobble}%
\providecommand \bibinfo  [0]{\@secondoftwo}%
\providecommand \bibfield  [0]{\@secondoftwo}%
\providecommand \translation [1]{[#1]}%
\providecommand \BibitemOpen [0]{}%
\providecommand \bibitemStop [0]{}%
\providecommand \bibitemNoStop [0]{.\EOS\space}%
\providecommand \EOS [0]{\spacefactor3000\relax}%
\providecommand \BibitemShut  [1]{\csname bibitem#1\endcsname}%
\let\auto@bib@innerbib\@empty
\bibitem [{\citenamefont {Strogatz}\ and\ \citenamefont
  {Stewart}(1993)}]{StrogatzStewart}%
  \BibitemOpen
  \bibfield  {author} {\bibinfo {author} {\bibfnamefont {S.~H.}\ \bibnamefont
  {Strogatz}}\ and\ \bibinfo {author} {\bibfnamefont {I.}~\bibnamefont
  {Stewart}},\ }\href@noop {} {\bibfield  {journal} {\bibinfo  {journal}
  {Scientific American}\ }\textbf {\bibinfo {volume} {269}},\ \bibinfo {pages}
  {102} (\bibinfo {year} {1993})}\BibitemShut {NoStop}%
\bibitem [{\citenamefont {Kuramoto}(1984)}]{Kuramoto}%
  \BibitemOpen
  \bibfield  {author} {\bibinfo {author} {\bibfnamefont {Y.}~\bibnamefont
  {Kuramoto}},\ }\href@noop {} {\emph {\bibinfo {title} {Chemical Oscillations,
  Waves, and Turbulence.}}}\ (\bibinfo  {publisher} {Springer-Verlag},\
  \bibinfo {address} {Berlin},\ \bibinfo {year} {1984})\BibitemShut {NoStop}%
\bibitem [{\citenamefont {Acebr{\'o}n}\ \emph {et~al.}(2005)\citenamefont
  {Acebr{\'o}n}, \citenamefont {Bonilla}, \citenamefont {Vicente},
  \citenamefont {Ritort},\ and\ \citenamefont {Spigler}}]{Acebron}%
  \BibitemOpen
  \bibfield  {author} {\bibinfo {author} {\bibfnamefont {J.}~\bibnamefont
  {Acebr{\'o}n}}, \bibinfo {author} {\bibfnamefont {L.}~\bibnamefont
  {Bonilla}}, \bibinfo {author} {\bibfnamefont {C.~P.}\ \bibnamefont
  {Vicente}}, \bibinfo {author} {\bibfnamefont {F.}~\bibnamefont {Ritort}}, \
  and\ \bibinfo {author} {\bibfnamefont {R.}~\bibnamefont {Spigler}},\
  }\href@noop {} {\bibfield  {journal} {\bibinfo  {journal} {Rev. Mod. Phys.}\
  }\textbf {\bibinfo {volume} {77}},\ \bibinfo {pages} {137} (\bibinfo {year}
  {2005})}\BibitemShut {NoStop}%
\bibitem [{\citenamefont {Lewis}(2003)}]{Lewis}%
  \BibitemOpen
  \bibfield  {author} {\bibinfo {author} {\bibfnamefont {J.}~\bibnamefont
  {Lewis}},\ }\href@noop {} {\bibfield  {journal} {\bibinfo  {journal} {Curr.
  Biol.}\ }\textbf {\bibinfo {volume} {13}},\ \bibinfo {pages} {1398} (\bibinfo
  {year} {2003})}\BibitemShut {NoStop}%
\bibitem [{\citenamefont {Garcia-Ojalvo}\ \emph {et~al.}(2004)\citenamefont
  {Garcia-Ojalvo}, \citenamefont {Elowitz},\ and\ \citenamefont
  {Strogatz}}]{GarciaOjalvo}%
  \BibitemOpen
  \bibfield  {author} {\bibinfo {author} {\bibfnamefont {J.}~\bibnamefont
  {Garcia-Ojalvo}}, \bibinfo {author} {\bibfnamefont {M.~B.}\ \bibnamefont
  {Elowitz}}, \ and\ \bibinfo {author} {\bibfnamefont {S.~H.}\ \bibnamefont
  {Strogatz}},\ }\href@noop {} {\bibfield  {journal} {\bibinfo  {journal}
  {Proc.~Natl.~Acad.~Sci.~USA}\ }\textbf {\bibinfo {volume} {101}},\ \bibinfo
  {pages} {10955} (\bibinfo {year} {2004})}\BibitemShut {NoStop}%
\bibitem [{\citenamefont {Yeung}\ and\ \citenamefont {Strogatz}(1999)}]{Yeung}%
  \BibitemOpen
  \bibfield  {author} {\bibinfo {author} {\bibfnamefont {M.~K.~S.}\
  \bibnamefont {Yeung}}\ and\ \bibinfo {author} {\bibfnamefont {S.~H.}\
  \bibnamefont {Strogatz}},\ }\href@noop {} {\bibfield  {journal} {\bibinfo
  {journal} {Phys. Rev. Lett.}\ }\textbf {\bibinfo {volume} {82}},\ \bibinfo
  {pages} {648} (\bibinfo {year} {1999})}\BibitemShut {NoStop}%
\bibitem [{\citenamefont {Izhikevich}(1998)}]{Izhikevich}%
  \BibitemOpen
  \bibfield  {author} {\bibinfo {author} {\bibfnamefont {E.~M.}\ \bibnamefont
  {Izhikevich}},\ }\href {\doibase 10.1103/PhysRevE.58.905} {\bibfield
  {journal} {\bibinfo  {journal} {Phys. Rev. E}\ }\textbf {\bibinfo {volume}
  {58}},\ \bibinfo {pages} {905} (\bibinfo {year} {1998})}\BibitemShut
  {NoStop}%
\bibitem [{\citenamefont {Ares}\ \emph {et~al.}(2012)\citenamefont {Ares},
  \citenamefont {Morelli}, \citenamefont {J\"org}, \citenamefont {Oates},\ and\
  \citenamefont {J\"ulicher}}]{Ares}%
  \BibitemOpen
  \bibfield  {author} {\bibinfo {author} {\bibfnamefont {S.}~\bibnamefont
  {Ares}}, \bibinfo {author} {\bibfnamefont {L.~G.}\ \bibnamefont {Morelli}},
  \bibinfo {author} {\bibfnamefont {D.~J.}\ \bibnamefont {J\"org}}, \bibinfo
  {author} {\bibfnamefont {A.~C.}\ \bibnamefont {Oates}}, \ and\ \bibinfo
  {author} {\bibfnamefont {F.}~\bibnamefont {J\"ulicher}},\ }\href@noop {}
  {\bibfield  {journal} {\bibinfo  {journal} {Phys. Rev. Lett.}\ }\textbf
  {\bibinfo {volume} {108}},\ \bibinfo {pages} {204101} (\bibinfo {year}
  {2012})}\BibitemShut {NoStop}%
\bibitem [{\citenamefont {Herrgen}\ \emph {et~al.}(2010)\citenamefont
  {Herrgen}, \citenamefont {Ares}, \citenamefont {Morelli}, \citenamefont
  {Schr{\"o}ter}, \citenamefont {J{\"u}licher},\ and\ \citenamefont
  {Oates}}]{Herrgen}%
  \BibitemOpen
  \bibfield  {author} {\bibinfo {author} {\bibfnamefont {L.}~\bibnamefont
  {Herrgen}}, \bibinfo {author} {\bibfnamefont {S.}~\bibnamefont {Ares}},
  \bibinfo {author} {\bibfnamefont {L.~G.}\ \bibnamefont {Morelli}}, \bibinfo
  {author} {\bibfnamefont {C.}~\bibnamefont {Schr{\"o}ter}}, \bibinfo {author}
  {\bibfnamefont {F.}~\bibnamefont {J{\"u}licher}}, \ and\ \bibinfo {author}
  {\bibfnamefont {A.~C.}\ \bibnamefont {Oates}},\ }\href@noop {} {\bibfield
  {journal} {\bibinfo  {journal} {Curr. Biol.}\ }\textbf {\bibinfo {volume}
  {20}},\ \bibinfo {pages} {1244 } (\bibinfo {year} {2010})}\BibitemShut
  {NoStop}%
\bibitem [{\citenamefont {Dhamala}\ \emph {et~al.}(2004)\citenamefont
  {Dhamala}, \citenamefont {Jirsa},\ and\ \citenamefont {Ding}}]{Dhamala}%
  \BibitemOpen
  \bibfield  {author} {\bibinfo {author} {\bibfnamefont {M.}~\bibnamefont
  {Dhamala}}, \bibinfo {author} {\bibfnamefont {V.~K.}\ \bibnamefont {Jirsa}},
  \ and\ \bibinfo {author} {\bibfnamefont {M.}~\bibnamefont {Ding}},\ }\href
  {\doibase 10.1103/PhysRevLett.92.074104} {\bibfield  {journal} {\bibinfo
  {journal} {Phys. Rev. Lett.}\ }\textbf {\bibinfo {volume} {92}},\ \bibinfo
  {pages} {074104} (\bibinfo {year} {2004})}\BibitemShut {NoStop}%
\bibitem [{\citenamefont {Rosenblum}\ and\ \citenamefont
  {Pikovsky}(2004)}]{Rosenblum}%
  \BibitemOpen
  \bibfield  {author} {\bibinfo {author} {\bibfnamefont {M.~G.}\ \bibnamefont
  {Rosenblum}}\ and\ \bibinfo {author} {\bibfnamefont {A.~S.}\ \bibnamefont
  {Pikovsky}},\ }\href@noop {} {\bibfield  {journal} {\bibinfo  {journal}
  {Phys. Rev. Lett.}\ }\textbf {\bibinfo {volume} {92}},\ \bibinfo {pages}
  {114102} (\bibinfo {year} {2004})}\BibitemShut {NoStop}%
\bibitem [{\citenamefont {Sakaguchi}\ and\ \citenamefont
  {Kuramoto}(1986)}]{SakaguchiKuramoto}%
  \BibitemOpen
  \bibfield  {author} {\bibinfo {author} {\bibfnamefont {H.}~\bibnamefont
  {Sakaguchi}}\ and\ \bibinfo {author} {\bibfnamefont {Y.}~\bibnamefont
  {Kuramoto}},\ }\href@noop {} {\bibfield  {journal} {\bibinfo  {journal}
  {Progress of Theoretical Physics}\ }\textbf {\bibinfo {volume} {76}},\
  \bibinfo {pages} {576} (\bibinfo {year} {1986})}\BibitemShut {NoStop}%
\bibitem [{\citenamefont {Schuster}\ and\ \citenamefont
  {Wagner}(1989)}]{SchusterWagner}%
  \BibitemOpen
  \bibfield  {author} {\bibinfo {author} {\bibfnamefont {H.}~\bibnamefont
  {Schuster}}\ and\ \bibinfo {author} {\bibfnamefont {P.}~\bibnamefont
  {Wagner}},\ }\href@noop {} {\bibfield  {journal} {\bibinfo  {journal} {Prog.
  Theor. Phys}\ }\textbf {\bibinfo {volume} {81}},\ \bibinfo {pages} {939}
  (\bibinfo {year} {1989})}\BibitemShut {NoStop}%
\bibitem [{\citenamefont {Wetzel}(2012)}]{Wetzel2012}%
  \BibitemOpen
  \bibfield  {author} {\bibinfo {author} {\bibfnamefont {L.}~\bibnamefont
  {Wetzel}},\ }\emph {\bibinfo {title} {Effect of Distributed Delays in Systems
  of Coupled Phase Oscillators}},\ \href@noop {} {Ph.D. thesis},\ \bibinfo
  {school} {TU Dresden} (\bibinfo {year} {2012})\BibitemShut {NoStop}%
\bibitem [{Note1()}]{Note1}%
  \BibitemOpen
  \bibinfo {note} {We prepare the system in the synchronized state and
  introduce small random perturbations to all phases. We let this perturbed
  state relax, and measure the Kuramoto order parameter $Z(t) = N^{-1} | \DOTSB
  \sum@ \slimits@ _i \protect \mathrm {e}^{\protect \mathrm {i}\theta _i(t)} |$
  as it approaches one. Relaxation becomes exponential for large times. We
  determine the relaxation rate from a fit to this exponential.}\BibitemShut
  {Stop}%
\bibitem [{\citenamefont {Earl}\ and\ \citenamefont
  {Strogatz}(2003)}]{EarlStrogatz}%
  \BibitemOpen
  \bibfield  {author} {\bibinfo {author} {\bibfnamefont {M.~G.}\ \bibnamefont
  {Earl}}\ and\ \bibinfo {author} {\bibfnamefont {S.~H.}\ \bibnamefont
  {Strogatz}},\ }\href {\doibase 10.1103/PhysRevE.67.036204} {\bibfield
  {journal} {\bibinfo  {journal} {Phys. Rev. E}\ }\textbf {\bibinfo {volume}
  {67}},\ \bibinfo {pages} {036204} (\bibinfo {year} {2003})}\BibitemShut
  {NoStop}%
\bibitem [{\citenamefont {MacDonald}(1989)}]{MacDonald}%
  \BibitemOpen
  \bibfield  {author} {\bibinfo {author} {\bibfnamefont {N.}~\bibnamefont
  {MacDonald}},\ }\href@noop {} {\emph {\bibinfo {title} {Biological delay
  systems: linear stability theory}}}\ (\bibinfo  {publisher} {Cambridge
  University Press},\ \bibinfo {year} {1989})\BibitemShut {NoStop}%
\bibitem [{\citenamefont {Atay}(2010)}]{Atay}%
  \BibitemOpen
  \bibinfo {editor} {\bibfnamefont {F.~M.}\ \bibnamefont {Atay}},\ ed.,\
  \href@noop {} {\emph {\bibinfo {title} {Complex Time-Delay Systems}}}\
  (\bibinfo  {publisher} {Springer-Verlag},\ \bibinfo {year}
  {2010})\BibitemShut {NoStop}%
\bibitem [{\citenamefont {Amann}\ \emph {et~al.}(2007)\citenamefont {Amann},
  \citenamefont {Sch\"oll},\ and\ \citenamefont {Just}}]{Amann}%
  \BibitemOpen
  \bibfield  {author} {\bibinfo {author} {\bibfnamefont {A.}~\bibnamefont
  {Amann}}, \bibinfo {author} {\bibfnamefont {E.}~\bibnamefont {Sch\"oll}}, \
  and\ \bibinfo {author} {\bibfnamefont {W.}~\bibnamefont {Just}},\ }\href@noop
  {} {\bibfield  {journal} {\bibinfo  {journal} {Physica A}\ }\textbf {\bibinfo
  {volume} {373}},\ \bibinfo {pages} {191} (\bibinfo {year}
  {2007})}\BibitemShut {NoStop}%
\bibitem [{\citenamefont {Corless}\ \emph {et~al.}(1996)\citenamefont
  {Corless}, \citenamefont {Gonnet}, \citenamefont {Hare}, \citenamefont
  {Jeffrey},\ and\ \citenamefont {Knuth}}]{Corless}%
  \BibitemOpen
  \bibfield  {author} {\bibinfo {author} {\bibfnamefont {R.}~\bibnamefont
  {Corless}}, \bibinfo {author} {\bibfnamefont {G.}~\bibnamefont {Gonnet}},
  \bibinfo {author} {\bibfnamefont {D.}~\bibnamefont {Hare}}, \bibinfo {author}
  {\bibfnamefont {D.}~\bibnamefont {Jeffrey}}, \ and\ \bibinfo {author}
  {\bibfnamefont {D.}~\bibnamefont {Knuth}},\ }\href@noop {} {\bibfield
  {journal} {\bibinfo  {journal} {Adv. Comput. Math.}\ }\textbf {\bibinfo
  {volume} {5}},\ \bibinfo {pages} {329} (\bibinfo {year} {1996})}\BibitemShut
  {NoStop}%
\bibitem [{\citenamefont {Shinozaki}\ and\ \citenamefont
  {Mori}(2006)}]{ShinozakiMori}%
  \BibitemOpen
  \bibfield  {author} {\bibinfo {author} {\bibfnamefont {H.}~\bibnamefont
  {Shinozaki}}\ and\ \bibinfo {author} {\bibfnamefont {T.}~\bibnamefont
  {Mori}},\ }\href {\doibase 10.1016/j.automatica.2006.05.008} {\bibfield
  {journal} {\bibinfo  {journal} {Automatica}\ }\textbf {\bibinfo {volume}
  {42}},\ \bibinfo {pages} {1791} (\bibinfo {year} {2006})}\BibitemShut
  {NoStop}%
\bibitem [{\citenamefont {Heagy}\ \emph {et~al.}(1995)\citenamefont {Heagy},
  \citenamefont {Pecora},\ and\ \citenamefont {Carroll}}]{Heagy}%
  \BibitemOpen
  \bibfield  {author} {\bibinfo {author} {\bibfnamefont {J.~F.}\ \bibnamefont
  {Heagy}}, \bibinfo {author} {\bibfnamefont {L.~M.}\ \bibnamefont {Pecora}}, \
  and\ \bibinfo {author} {\bibfnamefont {T.~L.}\ \bibnamefont {Carroll}},\
  }\href {http://prl.aps.org/abstract/PRL/v74/i21/p4185\_1} {\bibfield
  {journal} {\bibinfo  {journal} {Phys. Rev. Lett.}\ }\textbf {\bibinfo
  {volume} {74}},\ \bibinfo {pages} {4185} (\bibinfo {year}
  {1995})}\BibitemShut {NoStop}%
\bibitem [{\citenamefont {Pikovsky}\ \emph {et~al.}(2001)\citenamefont
  {Pikovsky}, \citenamefont {Rosenblum},\ and\ \citenamefont
  {Kurths}}]{Pikovsky}%
  \BibitemOpen
  \bibfield  {author} {\bibinfo {author} {\bibfnamefont {A.}~\bibnamefont
  {Pikovsky}}, \bibinfo {author} {\bibfnamefont {M.}~\bibnamefont {Rosenblum}},
  \ and\ \bibinfo {author} {\bibfnamefont {J.}~\bibnamefont {Kurths}},\
  }\href@noop {} {\emph {\bibinfo {title} {Synchronization. A universal concept
  in nonlinear sciences}}}\ (\bibinfo  {publisher} {Cambridge University
  Press},\ \bibinfo {address} {Cambridge},\ \bibinfo {year} {2001})\BibitemShut
  {NoStop}%
\bibitem [{\citenamefont {Manrubia}\ \emph {et~al.}(2004)\citenamefont
  {Manrubia}, \citenamefont {Mikhailov},\ and\ \citenamefont
  {Zanette}}]{Manrubia}%
  \BibitemOpen
  \bibfield  {author} {\bibinfo {author} {\bibfnamefont {S.}~\bibnamefont
  {Manrubia}}, \bibinfo {author} {\bibfnamefont {A.}~\bibnamefont {Mikhailov}},
  \ and\ \bibinfo {author} {\bibfnamefont {D.}~\bibnamefont {Zanette}},\
  }\href@noop {} {\emph {\bibinfo {title} {Emergence of dynamical order:
  synchronization phenomena in complex systems}}}\ (\bibinfo  {publisher}
  {World Scientific},\ \bibinfo {year} {2004})\BibitemShut {NoStop}%
\bibitem [{\citenamefont {Riedel-Kruse}\ \emph {et~al.}(2007)\citenamefont
  {Riedel-Kruse}, \citenamefont {M\"uller},\ and\ \citenamefont
  {Oates}}]{RiedelKruse}%
  \BibitemOpen
  \bibfield  {author} {\bibinfo {author} {\bibfnamefont {I.~H.}\ \bibnamefont
  {Riedel-Kruse}}, \bibinfo {author} {\bibfnamefont {C.}~\bibnamefont
  {M\"uller}}, \ and\ \bibinfo {author} {\bibfnamefont {A.~C.}\ \bibnamefont
  {Oates}},\ }\href@noop {} {\bibfield  {journal} {\bibinfo  {journal}
  {Science}\ }\textbf {\bibinfo {volume} {317}},\ \bibinfo {pages} {1911}
  (\bibinfo {year} {2007})}\BibitemShut {NoStop}%
\bibitem [{\citenamefont {Zhang}\ and\ \citenamefont {Kay}(2010)}]{Zhang}%
  \BibitemOpen
  \bibfield  {author} {\bibinfo {author} {\bibfnamefont {E.~E.}\ \bibnamefont
  {Zhang}}\ and\ \bibinfo {author} {\bibfnamefont {S.~A.}\ \bibnamefont
  {Kay}},\ }\href@noop {} {\bibfield  {journal} {\bibinfo  {journal} {Nat. Rev.
  Mol. Cell Biol.}\ }\textbf {\bibinfo {volume} {11}},\ \bibinfo {pages} {764}
  (\bibinfo {year} {2010})}\BibitemShut {NoStop}%
\bibitem [{\citenamefont {Schibler}\ and\ \citenamefont
  {Naef}(2005)}]{Schibler}%
  \BibitemOpen
  \bibfield  {author} {\bibinfo {author} {\bibfnamefont {U.}~\bibnamefont
  {Schibler}}\ and\ \bibinfo {author} {\bibfnamefont {F.}~\bibnamefont
  {Naef}},\ }\href@noop {} {\bibfield  {journal} {\bibinfo  {journal} {Curr.
  opin. cell biol.}\ }\textbf {\bibinfo {volume} {17}},\ \bibinfo {pages} {223}
  (\bibinfo {year} {2005})}\BibitemShut {NoStop}%
\bibitem [{\citenamefont {Gonze}(2011)}]{Gonze2}%
  \BibitemOpen
  \bibfield  {author} {\bibinfo {author} {\bibfnamefont {D.}~\bibnamefont
  {Gonze}},\ }\href@noop {} {\bibfield  {journal} {\bibinfo  {journal} {Cent.
  Eur. J. Biol.}\ }\textbf {\bibinfo {volume} {6}},\ \bibinfo {pages} {712}
  (\bibinfo {year} {2011})}\BibitemShut {NoStop}%
\bibitem [{\citenamefont {Oates}\ \emph {et~al.}(2012)\citenamefont {Oates},
  \citenamefont {Morelli},\ and\ \citenamefont {Ares}}]{Oates}%
  \BibitemOpen
  \bibfield  {author} {\bibinfo {author} {\bibfnamefont {A.~C.}\ \bibnamefont
  {Oates}}, \bibinfo {author} {\bibfnamefont {L.~G.}\ \bibnamefont {Morelli}},
  \ and\ \bibinfo {author} {\bibfnamefont {S.}~\bibnamefont {Ares}},\
  }\href@noop {} {\bibfield  {journal} {\bibinfo  {journal} {Development}\
  }\textbf {\bibinfo {volume} {139}},\ \bibinfo {pages} {625} (\bibinfo {year}
  {2012})}\BibitemShut {NoStop}%
\bibitem [{\citenamefont {Cooke}\ and\ \citenamefont {Zeeman}(1976)}]{Cooke}%
  \BibitemOpen
  \bibfield  {author} {\bibinfo {author} {\bibfnamefont {J.}~\bibnamefont
  {Cooke}}\ and\ \bibinfo {author} {\bibfnamefont {E.~C.}\ \bibnamefont
  {Zeeman}},\ }\href@noop {} {\bibfield  {journal} {\bibinfo  {journal} {J.
  Theor. Biol.}\ }\textbf {\bibinfo {volume} {58}},\ \bibinfo {pages} {455}
  (\bibinfo {year} {1976})}\BibitemShut {NoStop}%
\bibitem [{\citenamefont {Morelli}\ \emph {et~al.}(2009)\citenamefont
  {Morelli}, \citenamefont {Ares}, \citenamefont {Herrgen}, \citenamefont
  {Schr{\"o}ter}, \citenamefont {J{\"u}licher},\ and\ \citenamefont
  {Oates}}]{Morelli}%
  \BibitemOpen
  \bibfield  {author} {\bibinfo {author} {\bibfnamefont {L.~G.}\ \bibnamefont
  {Morelli}}, \bibinfo {author} {\bibfnamefont {S.}~\bibnamefont {Ares}},
  \bibinfo {author} {\bibfnamefont {L.}~\bibnamefont {Herrgen}}, \bibinfo
  {author} {\bibfnamefont {C.}~\bibnamefont {Schr{\"o}ter}}, \bibinfo {author}
  {\bibfnamefont {F.}~\bibnamefont {J{\"u}licher}}, \ and\ \bibinfo {author}
  {\bibfnamefont {A.~C.}\ \bibnamefont {Oates}},\ }\href@noop {} {\bibfield
  {journal} {\bibinfo  {journal} {HFSP J.}\ }\textbf {\bibinfo {volume} {3}},\
  \bibinfo {pages} {55} (\bibinfo {year} {2009})}\BibitemShut {NoStop}%
\bibitem [{\citenamefont {Kozyreff}\ \emph {et~al.}(2000)\citenamefont
  {Kozyreff}, \citenamefont {Vladimirov},\ and\ \citenamefont
  {Mandel}}]{Kozyreff}%
  \BibitemOpen
  \bibfield  {author} {\bibinfo {author} {\bibfnamefont {G.}~\bibnamefont
  {Kozyreff}}, \bibinfo {author} {\bibfnamefont {A.~G.}\ \bibnamefont
  {Vladimirov}}, \ and\ \bibinfo {author} {\bibfnamefont {P.}~\bibnamefont
  {Mandel}},\ }\href@noop {} {\bibfield  {journal} {\bibinfo  {journal} {Phys.
  Rev. Lett.}\ }\textbf {\bibinfo {volume} {85}},\ \bibinfo {pages} {3809}
  (\bibinfo {year} {2000})}\BibitemShut {NoStop}%
\bibitem [{\citenamefont {Tousi}\ \emph {et~al.}(2012)\citenamefont {Tousi},
  \citenamefont {Pourahmad},\ and\ \citenamefont {Afshari}}]{Tousi}%
  \BibitemOpen
  \bibfield  {author} {\bibinfo {author} {\bibfnamefont {Y.~M.}\ \bibnamefont
  {Tousi}}, \bibinfo {author} {\bibfnamefont {V.}~\bibnamefont {Pourahmad}}, \
  and\ \bibinfo {author} {\bibfnamefont {E.}~\bibnamefont {Afshari}},\
  }\href@noop {} {\bibfield  {journal} {\bibinfo  {journal} {Phys. Rev. Lett.}\
  }\textbf {\bibinfo {volume} {108}},\ \bibinfo {pages} {234101} (\bibinfo
  {year} {2012})}\BibitemShut {NoStop}%
\end{thebibliography}
\end{document}